\documentclass[usenatbib,a4paper,fleqn,times]{mn2e}
\usepackage{natbib,times,url}
\usepackage{aas_macros}	   
\usepackage{amsmath}       
\usepackage{amssymb}
\usepackage{wasysym}	   
\usepackage{graphicx}      
\usepackage{setspace}      
\usepackage{epstopdf}      
\usepackage{bm}			   
\usepackage{fleqn}		   
\usepackage{verbatim}

\title[Ambipolar diffusion in SPMHD]{Ambipolar diffusion in smoothed particle magnetohydrodynamics}

\author[Wurster, Price \& Ayliffe]{James Wurster\thanks{james.wurster@monash.edu}, Daniel Price\thanks{daniel.price@monash.edu} and Ben Ayliffe \\
Monash Centre for Astrophysics (MoCA), School of Mathematical Sciences, Monash University, Vic. 3800, Australia 
}
\date{Submitted: Revised: Accepted: }
\pagerange{\pageref{firstpage}--\pageref{lastpage}} \pubyear{2014}

\begin{document}
\label{firstpage}
\bibliographystyle{mn2e}
\maketitle

\begin{abstract}
In partially ionised plasmas, the magnetic field can become decoupled from the neutral gas and diffuse through it in a process known as ambipolar diffusion.  Although ambipolar diffusion has been implemented in several grid codes, we here provide an implementation in smoothed particle magnetohydrodynamics (SPMHD).  We use the strong coupling approximation in which the ion density is negligible, allowing a single fluid approach. The equations are derived to conserve energy, and to provide a positive definite contribution to the entropy. We test the implementation in both a simple 1D SPMHD code and the fully 3D code {\sc Phantom}. The wave damping test yields agreement within 0.03-2\% of the analytical result, depending on the value of the collisional coupling constant.  The oblique C-shocks test yields results that typically agree within 4\% of the semi-analytical result.  Our algorithm is therefore suitable for exploring the effect ambipolar diffusion has on physical processes, such as the formation of stars from molecular clouds. 
\end{abstract}


\section{Introduction}
 The ionisation fraction in molecular clouds is low ($\sim$$~10^{-16}$--$10^{-6}$). This means that ideal magnetohydrodynamics (MHD) is a poor approximation for magnetic fields in the star formation process. Non-ideal MHD terms arising from collisional multi-fluid interactions, including resistivity, ambipolar (ion-neutral) diffusion and the Hall effect all become important \citep{wardleng99,pgb08,PW2008}. Indeed, ambipolar diffusion formed the basis for the `standard model' of star formation \citep{mestelspitzer56,mouschoviaspaleologou81,sal87} that remains the subject of contentious debate.
 
  Most implementations of MHD into the smoothed particle hydrodynamics numerical method, referred to as smoothed particle magnetohydrodynamics (SPMHD; see review by \citealt{price12}) have considered only ideal MHD. This is because the primary difficulty in all numerical MHD implementations is the enforcement of the $\nabla\cdot \bm{B} = 0$ constraint. This led to the use of the Euler potentials, proposed earlier by \citet{phillipsmonaghan85}, where the magnetic field is written in the form $\bm{B} = \nabla\alpha \times \nabla\beta$ \citep{stern66}, enforcing the divergence constraint by construction. This was implemented by \citet{pricebate07} and \citet{rosswogprice07}, and subsequently used to study star formation problems \citep{pricebate07,pricebate08,pricebate09}. However, the Euler potentials cannot represent general 3D field geometries (since the magnetic Helicity $\bm{A}\cdot\bm{B} \equiv 0$), meaning that dynamo processes could not be simulated. Also, a physically consistent implementation of non-ideal MHD using the Euler potentials is difficult, if not impossible \citep{brandenburg10}. This led \citet{price10} to investigate a formulation based on the vector potential, $\bm{B} = \nabla\times\bm{A}$. Incorporation of non-ideal terms is then straightforward, but the SPMHD method based on the vector potential was found to be unstable, rendering this approach unusable.

 More recently, \citet{burzleetal11,burzleetal11a} demonstrated that the standard approach to SPMHD, where $\bm{B}$ or $\bm{B}/\rho$ is evolved \citep[c.f.][]{pricemonaghan04,pricemonaghan04a,pricemonaghan05} could indeed be used for a limited range of problems in star formation by using the artificial resistivity term to control divergence errors. This motivated \citet{triccoprice12} to re-investigate the application of \citet{dedneretal02}'s hyperbolic/parabolic divergence cleaning scheme to SPMHD. When implemented in a conservative form, this was found to reduce divergence errors by more than an order of magnitude compared to artificial resistivity alone, and was found to be at least as effective as the Euler potentials. This provides an SPMHD algorithm for star formation that is robust and general, which has already been used to simulate jets and outflows from protostars \citep{ptb12,btp14} and turbulent dynamos \citep*{tpf14}, and which allows non-ideal MHD terms to be incorporated in a straightforward manner. 

 The simplest non-ideal MHD effect, Ohmic resistivity, has already been implemented into SPMHD by several authors \citep{bonafedeetal11,tii13}, since it is a straightforward implementation of a diffusion term in SPH for which standard formulations exist \citep{brookshaw85,clearymonaghan99}. The other two non-ideal MHD effects, ambipolar diffusion and the Hall effect, are more complicated. \citet{hoskingwhitworth04} were the first to model ambipolar diffusion in SPMHD, which they applied to star formation \citep{hoskingwhitworth04a}. However, their formulation was complicated by their attempt to model the ions and neutrals as two distinct sets of SPH particles, despite making the assumptions ($\rho_\text{ion} \ll \rho_\text{n}$ and zero ion pressure) that mean the mixture is adequately described as a single fluid mixture with additional diffusion terms in the evolution equation for the magnetic field \citep{pgb08,PW2008}. This is the most widely adopted approach to implementing ambipolar diffusion in Eulerian codes\footnote{We have recently made a similar argument regarding the usefulness of a single fluid description in the context of dust-gas mixtures \citep{laibeprice14}} (e.g. \citealt{MNKW95,DP08,CKW09,massonetal12}, though see \citealt{tilleybalsara11,tbm12}).
 
  In this paper we derive a much simpler algorithm for ambipolar diffusion in SPMHD based on the single fluid description, laying also the foundation for an implementation of the Hall effect. The paper is organised as follows: We discuss the continuum equations in Sec.~\ref{sec:CE}. The corresponding discrete equations are derived in Sec.~\ref{sec:numericsI}. In Sec.~\ref{sec:tests} we benchmark the algorithm on the standard test problems for ambipolar diffusion, summarising in Sec.~\ref{sec:summary}.
 
\section{Continuum equations}
\label{sec:CE}

In a partially ionised plasma, the ion pressure and momentum can be assumed to be negligible compared to the pressure and momentum of the neutrals; this is the strong coupling approximation.  In this case, the plasma can be treated as a single fluid, where the plasma density is equated to the neutral density $\rho \sim \rho_\text{n}$, and the ion density is small, $\rho_\text{ion} \ll \rho_\text{n}$.  Using this approximation, the MHD equations that describe the ionised plasma are
\begin{equation}
\label{CErho}
\frac{\text{d} \rho}{\text{d} t} =  -\rho \bm{\nabla} \cdot \bm{v},
\end{equation}
\begin{equation}
\label{CEv}
\frac{\text{d} \bm{v}}{\text{d} t} = -\frac{1}{\rho}\bm{\nabla} \left[\left(P+\frac{B^2}{2\mu_0}\right)I - \frac{\bm{B}\bm{B}}{\mu_0}\right],
\end{equation}
\begin{equation}
\label{CEu}
\frac{\text{d} u}{\text{d} t} = -\frac{P}{\rho} \bm{\nabla} \cdot \bm{v} + \left.\frac{\text{d} u}{\text{d} t}\right|_\text{AD},
\end{equation}
\begin{equation}
\label{CEb}
\frac{\text{d} \bm{B}}{\text{d} t} = \left(\bm{B}\cdot\bm{\nabla}\right)\bm{v}-\bm{B}\left(\bm{\nabla}\cdot\bm{v}\right) + \left.\frac{\text{d} \bm{B}}{\text{d} t}\right|_\text{AD},
\end{equation}
where $\frac{\text{d}}{\text{d}t} \equiv \frac{\partial}{\partial t} + \bm{v}\cdot\bm{\nabla}$ is the Lagrangian derivative, $\bm{v}$, $P$ and $u$ are the neutral velocity, pressure and specific internal energy, respectively, $\bm{B}$ is the magnetic field, $\mu_0$ is the permeability of free space, and $I$ is the identity matrix.  The final term on the right hand side of \eqref{CEb} describes the ambipolar diffusion, while the final term in \eqref{CEu} is to maintain conservation of energy.  The MHD equations are closed by the equation of state,
\begin{equation}
\label{CEP}
P = \left(\gamma - 1\right)\rho u,
\end{equation}
where $\gamma$ is the adiabatic index, and constrained by
\begin{equation}
\label{CEdivB}
\bm{\nabla} \cdot \bm{B} = 0.
\end{equation}

For a perfectly neutral plama, $\left.\frac{\text{d} \bm{B}}{\text{d} t}\right|_\text{AD} = 0$ and $\left.\frac{\text{d} u}{\text{d} t}\right|_\text{AD} = 0$.  For the partially ionised plasma we present here, we have 
\begin{equation}
\label{CEbambitmp}
\left.\frac{\text{d} \bm{B}}{\text{d} t}\right|_\text{AD} = \bm{\nabla} \times \left\{ \frac{\eta_\text{AD}}{B^2}\left[\left(\bm{\nabla}\times\bm{B}\right)\times\bm{B}\right]\times\bm{B}\right\}.
\end{equation}
The ambipolar diffusion coefficient is defined by
\begin{equation}
\eta_\text{AD} \equiv \frac{v_\text{A}^2}{\gamma_\text{AD}\rho_\text{ion}} ,
\end{equation}
where $\gamma_\text{AD}$ is the collisional coupling constant between ions and neutrals due to ambipolar diffusion \citep[see][]{PW2008}, $\rho_\text{ion}$ is the ion density, and $v_\text{A} \equiv B/\sqrt{\mu_0\rho}$ is the Alfv{\'e}n speed.  We further define
\begin{equation}
\label{CEd}
\bm{D} \equiv  \frac{\eta_\text{AD}\mu_0}{B^2}\left(\bm{J}\times\bm{B}\right)\times\bm{B}, 
\end{equation}
where $\bm{J} = \frac{1}{\mu_0}\bm{\nabla}\times\bm{B}$ is the magnetic current density.  Now, \eqref{CEbambitmp} can be simplified to
\begin{equation}
\label{CEbambi}
\left.\frac{\text{d} \bm{B}}{\text{d} t}\right|_\text{AD} = \bm{\nabla} \times \bm{D}.
\end{equation}
Although we are focusing on ambipolar diffusion with $\bm{D}$ defined as in \eqref{CEd}, \eqref{CEbambi} is a general form that can be used for all non-ideal MHD terms.  For example, $\bm{D}$ can be defined as
\begin{equation}
\bm{D}_\text{Ohmic} \equiv -\frac{1}{\sigma}\bm{J} = -\eta_\text{O}\mu_0\bm{J},
\end{equation}
or
\begin{equation}
\bm{D}_\text{Hall} \equiv -\frac{1}{e n_e }\bm{J}\times\bm{B} = -\frac{\eta_\text{H}\mu_0}{|\bm{B}|}\bm{J}\times\bm{B},
\end{equation}
for Ohmic and Hall diffusivity, respectively (e.g. \citealp{PW2008}), where $\sigma$ is the conductivity, $e$ is the electron charge, and $n_e$ is the electron number density.  Given this form of the $\eta$-coefficients, the units of $\eta_\text{AD}$, $\eta_\text{O}$ and $\eta_\text{H}$ are all the same (area per unit time), thus aiding a more general discussion of non-ideal MHD.  Thus, the procedure that follows (and especially the procedure in \S \ref{sec:numericsI}) can be used for any of these terms; specific consideration will only be required once we substitute in the actual expression for $\bm{D}$.  

The ion velocity, $\bm{v}_\text{ion}$, can be recovered from the coupling to the neutral velocity, viz.
\begin{equation}
\label{CEvion}
\bm{v}_\text{ion} = \bm{v} + \frac{\eta_\text{AD}\mu_0}{B^2}\bm{J}\times\bm{B}.
\end{equation}
For ambipolar diffusion, the characteristic time-scale is
\begin{equation}
\label{CStau}
\tau_\text{AD} = \frac{1}{\gamma_\text{AD}\rho_\text{ion}}
\end{equation}
and the characteristic length-scale is
\begin{equation}
\label{CSl}
L_\text{AD} = \tau_\text{AD}v_\text{A}.
\end{equation}

In general one requires a method to determine the ionisation fraction in order to specify the ambipolar diffusion coefficient. For star formation problems, this depends on chemistry involving dust grains, with cosmic rays as the main source of ionisation. There exist numerous attempts to solve the chemical reaction schemes \citep[e.g.][]{umebayashinakano90,wardleng99,sanoetal00,nnu02} or to provide a simple parameterisation of the ionisation as a function of density and temperature \citep[e.g.][]{normanheyvaerts85,mim06,tii13,keithwardle14}. Perhaps the simplest approach is to assume that the ion density is a power-law function of the neutral density \citep{Elmegreen79}:
\begin{equation}
\label{CErhoion}
\rho_\text{ion} = \rho_\text{ion,0}\left(\frac{\rho_\text{n}}{\rho_\text{n,0}}\right)^\alpha.
\end{equation}
In the molecular cloud regime, $\alpha = 0$ provides a reasonable approximation (e.g. \citealt{Nakano84}), and this is the approximation used for the test problems in this paper. This means that $\tau_\text{AD}$ is a constant for all time, which will be useful in the numerical implementation of ambipolar diffusion (see \S \ref{sec:numericsI}).  Thus, $\tau_\text{AD} = \eta_\text{AD}/v_\text{A}^2$ will be a more useful quantity in the numerical derivation of the equations whereas $\eta_\text{AD}$ will be more useful in the continuum derivation, although the two quantities are interchangeable.

\subsection{Conservation of energy}
\label{CEce}
Total energy is given by
\begin{align}
 E &= \int_V \left(\frac{1}{2} \bm{v}^2 + u +\frac{\bm{B}^2}{2\mu_0\rho} \right) \rho \text{d}V.
\label{CEte}
\end{align}
Thus, to conserve energy, we require $\frac{\text{d}E}{\text{d}t} = 0$, where also
\begin{equation}
\label{CEde}
\frac{\text{d}E}{\text{d}t} = \int_V \left(\bm{v}\cdot\frac{\text{d}\bm{v}}{\text{d}t} + \frac{\text{d}u}{\text{d}t} + \frac{\bm{B}}{\mu_0\rho}\cdot\frac{\text{d}\bm{B}}{\text{d}t} - \frac{\bm{B}^2}{2\mu_0\rho^2}\frac{\text{d}\rho}{\text{d}t}\right) \rho \text{d}V.
\end{equation}
The standard components of the SPH equations already conserve energy \citep[see][]{pricemonaghan04a}, so we need to consider only the additional terms resulting from ambipolar diffusion, i.e.,
\begin{equation}
\label{CEdea1}
\frac{\text{d}E}{\text{d}t} = \int_V \left( \left.\frac{\text{d}u}{\text{d}t}\right|_\text{AD} + \frac{\bm{B}}{\mu_0\rho}\cdot\left.\frac{\text{d}\bm{B}}{\text{d}t}\right|_\text{AD} \right) \rho \text{d}V = 0.
\end{equation}
Therefore, 
\begin{equation}
\label{CEdea2}
\int_V \left.\frac{\text{d}u}{\text{d}t}\right|_\text{AD}\rho \text{d}V = -\int_V \frac{\bm{B}}{\mu_0\rho}\cdot\left.\frac{\text{d}\bm{B}}{\text{d}t}\right|_\text{AD} \rho \text{d}V.
\end{equation}
Using \eqref{CEbambi}, this can be expanded as
\begin{equation}
\label{CEduI}
\int_V \left.\frac{\text{d}u}{\text{d}t}\right|_\text{AD}\rho \text{d}V = -\frac{1}{\mu_0}\int_V \bm{B}\cdot \left(\bm{\nabla} \times \bm{D}\right) \text{d}V.
\end{equation}
Integrating by parts and using the divergence theorem yields
\begin{align}
\int_V \left.\frac{\text{d}u}{\text{d}t}\right|_\text{AD}\rho \text{d}V = &-\frac{1}{\mu_0}\int_V \bm{D}\cdot\left(\bm{\nabla}\times\bm{B}\right) \text{d}V \notag \\
   &- \frac{1}{\mu_0}\int_V \bm{\nabla}\cdot\left(\bm{D}\times\bm{B}\right) \text{d}V, \notag \\
= &-\frac{1}{\mu_0}\int_V \bm{D}\cdot\left(\bm{\nabla}\times\bm{B}\right) \text{d}V \notag \\
   &- \frac{1}{\mu_0}\oint_{S} (\bm{D} \times \bm{B}) \cdot {\rm d}\bm{S} \notag \\
= &-\int_V \bm{D}\cdot \bm{J} \text{d}V,
\end{align}
where the surface integral vanishes because the magnetic field tends to zero at infinity. Therefore,
\begin{equation}
\label{eq:CEdu}
\left.\frac{\text{d}u}{\text{d}t}\right|_\text{AD} = -\frac{\bm{D}\cdot\bm{J}}{\rho}.
\end{equation}
Replacing $\bm{D}$ with \eqref{CEd} and converting to tensor notation (where $\epsilon_{ijk}$ is the Levi-Civita tensor and $\delta_{ij}$ is the Kronecker delta) yields
\begin{align*}
\left.\frac{\text{d}u}{\text{d}t}\right|_\text{AD} 
 =& -\frac{\eta_\text{AD}\mu_0}{B^2\rho} \epsilon_{ijk} \epsilon_{juv} J_i J_u B_k B_v 
\end{align*}
Using the standard relation $\epsilon_{ijk} \epsilon_{juv} = \delta_{iv}\delta_{ku} - \delta_{iu}\delta_{kv}$, simplifying, and returning to vector notation yields
\begin{equation}
\left.\frac{\text{d}u}{\text{d}t}\right|_\text{AD} = -\frac{\eta_\text{AD}\mu_0}{B^2\rho} \left[ \left(\bm{J}\cdot\bm{B}\right)^2 - J^2B^2 \right].
\end{equation}
Analytically, $\bm{J}\cdot\bm{B} = 0$, 
thus we have
\begin{equation}
\left.\frac{\text{d}u}{\text{d}t}\right|_\text{AD} = \frac{\eta_\text{AD}\mu_0 J^2}{\rho}.
\end{equation}
From the above we see that ambipolar diffusion results in a positive definite contribution to the thermal energy, and hence the entropy. In the case of the Hall effect, (\ref{eq:CEdu}) gives
\begin{equation}
 \left.\frac{{\rm d}u}{{\rm d}t}\right\vert_\text{Hall} = -\frac{\bm{D}_\text{Hall}\cdot \bm{J}}{\rho} = 0,
\end{equation}
indicating that the Hall effect is not dissipative.  For completeness, the dissipation term due to Ohmic resistivity is given by 
\begin{equation}
 \left.\frac{{\rm d}u}{{\rm d}t}\right|_\text{Ohmic} = -\frac{\bm{D}_\text{Ohmic}\cdot \bm{J}}{\rho} = \frac{\eta_\text{O} \mu_0 J^2}{\rho},
\end{equation}
which is also positive definite.
\section{Implementation in SPMHD}
\label{sec:numericsI}

The numerical implementation of the MHD equations without ambipolar diffusion has been previously described in the literature \citep[see review by][]{price12}; these sources include a thorough description of dissipation terms \citep{pricemonaghan04,pricemonaghan05,triccoprice13} and cleaning of the $\bm{\nabla} \cdot \bm{B}$ term \citep{pricemonaghan05,triccoprice12}.  For completeness, we will rewrite \eqref{CErho}--\eqref{CEb} and the ambipolar diffusion-related terms in their SPH form, but refer the reader to the literature for discussion on shock dissipation and cleaning.  For this implementation, we assume units such that $\mu_0\equiv1$ (see \citealt{pricemonaghan04a}). 

\subsection{Discrete equations}
The SPMHD equations (momentarily neglecting ambipolar diffusion) for particle $a$ are given by
\begin{equation}
\label{Irho}
\rho_a = \sum_b m_b W_{ab}(h_a),
\end{equation}
\begin{equation}
\label{Iv}
\frac{\text{d} v^i_a}{\text{d} t} = \sum_b m_b \left[\frac{S^{ij}_a}{\Omega_a \rho_a^2} \nabla^j_a W_{ab}(h_a) + \frac{S^{ij}_b}{\Omega_b \rho_b^2}\nabla^j_a W_{ab}(h_b)\right],
\end{equation}
\begin{equation}
\label{Iu}
\frac{\text{d} u_a}{\text{d} t} = \frac{P_a}{\Omega_a \rho_a^2} \sum_b m_b v^i_{ab} \nabla^i_a W_{ab}(h_a),
\end{equation}
\begin{align}
\label{Ib}
\frac{\text{d} B^i_a}{\text{d} t} = -\frac{1}{\Omega_a \rho_a} \sum_b m_b & \left[  v^i_{ab} B^j_a \nabla^j_a W_{ab}\left(h_a\right) \right. \notag \\
 - & \left. B^i_a v^j_{ab} \nabla^j_a W_{ab}\left(h_a\right) \right],
\end{align}
where we sum over all particles, $b$, within the kernel radius, $W_{ab}$ is the smoothing kernel, $\bm{v}_{ab} = \bm{v}_a - \bm{v}_b$, $\Omega_a$ is a dimensionless correction term to account for a spatially variable smoothing length \citep{monaghan02,springelhernquist02}, and the stress tensor is given by
\begin{equation}
\label{Istress}
S^{ij} \equiv -\left(P + \frac{1}{2}B^2\right)\delta^{ij} + B^iB^j.
\end{equation}

To calculate $\left.\frac{\text{d} \bm{B}}{\text{d} t}\right|_\text{AD}$, we first calculate the current density, $\bm{J} \equiv \bm{\nabla}\times\bm{B}$, using the difference operator (c.f. \citealt{price10,price12}):
\begin{align}
\label{IJ}
\bm{J}_a = \frac{1}{\Omega_a \rho_a} \sum_b m_b \left(\bm{B}_a - \bm{B}_b\right) \times \bm{\nabla}_a W_{ab}(h_a).
\end{align}
We note that we could have instead chosen the symmetric operator, since, at this stage, the form of $\bm{J}$ is arbitrary.  However, the form of the remaining equations is directly dependent on this choice, as we will show below.

Once $\bm{J}$ is calculated, $\bm{D}$ is then calculated via
\begin{equation}
\label{ID}
D^i_a = \frac{\tau_{AD}}{\rho_a}\epsilon_{ijk}\epsilon_{juv} J_a^u B_a^v B_a^k.
\end{equation}
In order to give a positive definite contribution to the entropy, we find that the conjugate curl operator must be used for the $\nabla\times\bm{D}$ term in $\left.\frac{\text{d} \bm{B}}{\text{d} t}\right|_\text{AD}$:
\begin{align}
\label{Injcbcb}
\left.\frac{\text{d} \bm{B}_a}{\text{d} t}\right|_\text{AD} = -\rho_a \sum_b m_b  & \left[ \frac{\bm{D}_a}{\Omega_a \rho^2_a}\times \bm{\nabla}_aW_{ab}(h_a)\right. \\ \notag
+ & \left. \frac{\bm{D}_b}{\Omega_b \rho^2_b}\times \bm{\nabla}_aW_{ab}(h_b) \right].
\end{align}
The need to use conjugate operators to conserve energy in SPH has been discussed by a number of authors \citep[e.g.][]{cumminsrudman99,price10,triccoprice12,price12} and is demonstrated in \S\ref{sec:Nce} below.

To numerically calculate the ion velocity, we simply use
\begin{equation}
\label{Ivion}
\bm{v}_{a,\text{ion}} = \bm{v}_a + \frac{\tau_\text{AD}}{\rho_a}\left(\bm{J} \times \bm{B}\right)_a.
\end{equation}

\subsection{Timestepping}
\label{sec:dt}
The Courant limited timestep is derived assuming ideal MHD.  By introducing non-ideal MHD terms into the equations, a new timestep must also be considered to ensure numerical stability.  In general, this is given by
\begin{equation}
\label{Idtgeneral}
\text{d}t_{a} \le \frac{h_a^2}{\eta},
\end{equation}
where $\eta \in \{\eta_\text{O}, \eta_\text{H},\eta_\text{AD}\}$.  For ambipolar diffusion, the timestep is 
\begin{equation}
\label{Idt}
\text{d}t_{\text{AD},a} \le  \frac{h_a^2}{\eta_\text{AD}} = \frac{h_a^2}{\tau_\text{AD} v_{\text{A},a}^2}.
\end{equation}
The dependence on $h^2$ can provide a strict constraint on timestepping; specifically, this timestep will become important when $h/\tau_\text{AD} \ll v_\text{A}$.  To avoid a globally small timestep (in scenarios where it is necessary), several options have been employed, including a ``heavy-ion'' approximation \citep{LiMckeeKlein06}, setting the ambipolar diffusion rate to zero below a given density threshhold \citep{NakamuraLi08} and super time-stepping (e.g. \citealt{AlexiadesEtAl96}; \citealt{OsullivanDownes06}; \citealt{CKW09}).  For simplicity, we use \eqref{Idt} as our ambipolar diffusion-limited timestep, and we will further comment on the timestep when we discuss our numerical results.  

\subsection{Conservation of energy}
\label{sec:Nce}
Similar to \S \ref{CEce}, we will calculate the numerical implementation of $\left.\frac{\text{d} u_a}{\text{d} t}\right|_\text{AD}$ using the conservation of energy and demonstrate that the resulting contribution to the thermal energy, and hence the entropy, is positive definite.  The SPH form of \eqref{CEdea1} is
\begin{equation}
\label{IceduAD}
\frac{\text{d} E}{\text{d} t} = \sum_a m_a \left[ \left.\frac{\text{d} u_a}{\text{d} t}\right|_\text{AD} + \frac{\bm{B}_a}{\rho_a} \cdot \left.\frac{\text{d} \bm{B}_a}{\text{d} t}\right|_\text{AD} \right] = 0.
\end{equation}
Therefore, to conserve energy, we must have
\begin{equation}
\sum_a m_a \left.\frac{\text{d} u_a}{\text{d} t}\right|_\text{AD}  = -\sum_a m_a \frac{\bm{B}_a}{\rho_a} \cdot \left.\frac{\text{d} \bm{B}_a}{\text{d} t}\right|_\text{AD}.
\end{equation}
Substituting in \eqref{Injcbcb} and rearranging the terms yields
\begin{align}
\sum_a m_a \left.\frac{\text{d} u_a}{\text{d} t}\right|_\text{AD}  &= -\sum_a m_a \frac{\bm{B}_a}{\rho_a}  \cdot \left\{ -\rho_a \sum_b m_b \right. \notag \\
 &  \left.\left[ \frac{\bm{D}_a}{\Omega_a \rho^2_a}\times \bm{\nabla}_aW_{ab}(h_a) + \frac{\bm{D}_b}{\Omega_b \rho^2_b}\times \bm{\nabla}_aW_{ab}(h_b) \right] \right\}, \notag \\
= &\sum_a \sum_b m_a m_b \bm{B}_a \cdot \left[\frac{\bm{D}_a}{\Omega_a \rho^2_a}\times \bm{\nabla}_a W_{ab}(h_a)\right] \notag \\
+ &\sum_b \sum_a m_b m_a \bm{B}_b \cdot \left[\frac{\bm{D}_a}{\Omega_a \rho^2_a}\times \bm{\nabla}_b W_{ba}(h_a)\right], \notag \\
=-&\sum_a \sum_b m_a m_b \frac{\bm{D}_a}{\Omega_a\rho_a^2} \cdot \left[\left(\bm{B}_a - \bm{B}_b \right) \times \bm{\nabla}_a W_{ab}(h_a)\right],
\end{align}
where in the second step we rearranged the summation indices, and in the third step used the identity $\bm{\nabla}_b W_{ab} = -\bm{\nabla}_a W_{ab}$ and the triple scalar product $\bm{B} \cdot \left(\bm{D}\times\bm{\nabla}W\right) = -\bm{D} \cdot \left(\bm{B}\times\bm{\nabla}W\right)$ to simplify the expression. Removing the sum over $a$ yields
\begin{equation}
\left.\frac{\text{d} u_a}{\text{d} t}\right|_\text{AD}  
= - \sum_b m_b \frac{\bm{D}_a}{\Omega_a\rho_a^2} \cdot \left[\left(\bm{B}_a - \bm{B}_b \right) \times \bm{\nabla}_a W_{ab}(h_a)\right], 
\end{equation}
where 
\begin{equation*}
\frac{1}{\Omega_a \rho_a} \sum_b m_b \left(\bm{B}_a - \bm{B}_b\right) \times \bm{\nabla}_a W_{ab}(h_a) \equiv \bm{J}_a^\text{diff},
\end{equation*}
is the current density using the differenced operator, as given in \eqref{IJ}. Therefore, this can be rewritten as
\begin{equation}
\left.\frac{\text{d} u_a}{\text{d} t}\right|_\text{AD}  = - \frac{\bm{D}_a}{\rho_a}\cdot \bm{J}_a^\text{diff}.
\label{eq:ddotJ}
\end{equation}
From (\ref{eq:ddotJ}) we can see that the contribution to the thermal energy will only be positive definite if the numerical operator used to compute $\bm{J}$ in (\ref{CEd}) is the same as the operator that appears in (\ref{eq:ddotJ}), i.e. $\bm{J}^\text{diff}$, which is indeed the case in our formulation (cf. Eq.~\ref{IJ}). Specifically, following the procedure in the continuum case yields
\begin{align}
\left.\frac{\text{d}u_a}{\text{d}t}\right|_\text{AD} 
 =& -\frac{\tau_\text{AD}}{\rho_a^2} \epsilon^{ijk} \epsilon^{juv} J_a^i J_a^u B_a^k B_a^v, \notag \\
 =& -\frac{\tau_\text{AD}}{\rho_a^2} \left(\delta^{iv}\delta^{ku} - \delta^{iu}\delta^{kv}\right) J_a^i J_a^u B_a^k B_a^v, \notag \\
 =& -\frac{\tau_\text{AD}}{\rho_a^2} \left(J_a^i J_a^k B_a^k B_a^i - J_a^i J_a^i B_a^k B_a^k \right).
\end{align}
Therefore, 
\begin{equation}
\left.\frac{\text{d}u_a}{\text{d}t}\right|_\text{AD} = \frac{\tau_\text{AD}}{\rho^2_a} \left[ \left(J_a^\text{diff}\right)^2\left(B_a\right)^2 - \left(\bm{J}_a^\text{diff}\cdot\bm{B}_a\right)^2 \right],
\end{equation}
which is positive definite. Unlike in the continuum case, we cannot be guaranteed that $\bm{J}_a^\text{diff}\cdot\bm{B}_a = 0$, so this term must be included to conserve energy.

The same procedure would apply to a numerical implementation of the Hall effect, in which case \eqref{eq:ddotJ} becomes 
\begin{equation}
\left.\frac{\text{d} u_a}{\text{d} t}\right|_\text{Hall}  = - \frac{\bm{D}^\text{Hall}_a}{\rho_a}\cdot \bm{J}_a^\text{diff},
\label{eq:ddotHall}
\end{equation}
This term will be zero if we use $\bm{J}^\text{diff}$ to compute $\bm{D}^\text{Hall}$ since $ \left( \bm{J}^\text{diff} \times \bm{B}\right)\cdot \bm{J}^\text{diff} = 0$.   Therefore, the implementation of the Hall effect is a straightforward extension to the procedure we describe here to implement ambipolar diffusion.  The complication is that Hall MHD contains a very fast wave, the Whistler mode \citep{PW2008}, that severely limits the timestep, meaning that an implicit timestepping scheme is required. 

In principle, resistivity can also be implemented following the same method, where
\begin{equation}
\left.\frac{\text{d} u_a}{\text{d} t}\right|_\text{Ohmic}  = - \frac{\bm{D}^\text{Ohmic}_a}{\rho_a}\cdot \bm{J}_a^\text{diff} = \frac{\eta_\text{O} \left(J^\text{diff}_a\right)^2}{\rho_a},
\label{eq:ddotOhm}
\end{equation}
which is also positive definite.  The considerations above are similar to the implementation of resistivity with the vector potential \citep{price10}.  However, the resistivity term is simple enough that direct second derivatives can be taken, which is the approach taken by \citet{bonafedeetal11} and \citet{tii13}.  It is not clear whether there is any advantage to a two first derivatives approach in this case.  As resistivity is not the focus of this paper, we do not investigate this matter further.  

\section{Numerical tests}
\label{sec:tests}
We test our ambipolar diffusion algorithm using an isothermal equation of state, $P = c_\text{s}^2\rho$.  Our primary tests are performed using a simple 1D SPMHD code, but we have also implemented the algorithm in the 3D code {\sc Phantom}, which we also test. The 1D code uses the equations described in Sec.~\ref{sec:numericsI}, the $M_{6}$ quintic smoothing kernel, and a term to correct for $\bm{\nabla} \cdot \bm{B} = 0$.  Our two tests are the wave damping test, and the C-shock, as described below.  

\subsection{Wave damping test}
To test ambipolar diffusion in the strong coupling approximation, we follow the evolution of a standing wave as done in \citet{CKW09}.  The dispersion relation for Alfv{\'e}n waves is \citep{B96}
\begin{equation}
\label{NWq}
\omega^2 + \frac{v_\text{A}^2 k^2}{\gamma_\text{AD}\rho_\text{i}} \omega i - v_\text{A}^2 k^2 = 0,
\end{equation}
where $\omega = \omega_\text{R} + \omega_\text{I}i$ is the complex angular frequency of the wave and $k$ is a wavenumber; in this notation, $\omega_\text{R}$, $\omega_\text{I}$ and $k$ are all explicitly real.  The solution to \eqref{NWq} is a damped oscillation, where the time dependence of the first normal mode is given by
\begin{equation}
\label{NWh}
h(t) = h_0 \left|\sin\left(\omega_\text{R} t\right)\right| e^{\omega_\text{I}t},
\end{equation}
where $h_0$ is the initial amplitude of the wave.  Alfv{\'e}n waves will always propagate for $\omega_\text{R} \ne 0$, however
ambipolar diffusion can prevent their propagation in a partially ionised medium when $k~>~2\gamma_\text{AD}\rho_\text{i}/v_\text{A}$ \citep{KulsrudPearce69}.

To test the decay of Alfv{\'e}n waves, we turn off all dissipation terms, set $\rho = 1$, $c_\text{s} = 1$, $\bm{B} = B_0\bm\hat{x}$ with $B_0 = 1$, $L = 1.0$ with periodic boundary conditions and $\rho_\text{i} = 0.1$.  We initialise the velocity to be a standing wave in the $z$-direction, viz.
\begin{equation}
\label{NWvz}
v_z(x) = v_0 \sin(kx),
\end{equation}
where $k = 2\pi/L$ and $v_0 = 0.01v_\text{A}$.  We then track the oscillation of the root-mean-square magnetic field in the $z$-direction, $<B_z^2>^{1/2}$.  For this quantity, the initial amplitude in \eqref{NWh} is given by $h_0 = \frac{v_0}{v_\text{A}}\frac{B_0}{\sqrt{2}}$.  

\begin{figure}
\begin{center}
\includegraphics[width=1.0\columnwidth]{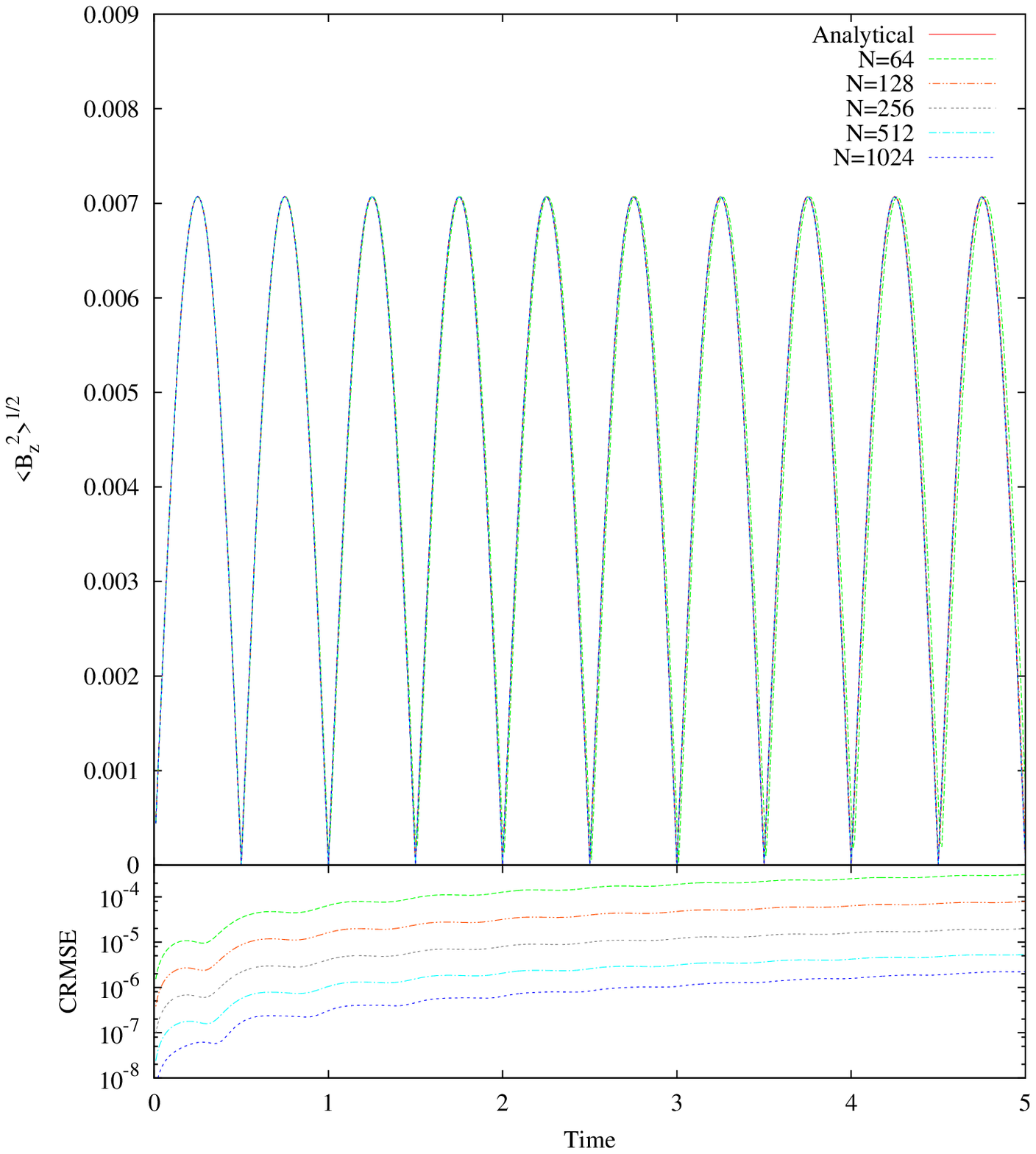}
\end{center}
\caption{\emph{Top}: Time evolution of the spatially averaged root-mean-square magnetic field in the $z$-direction without ambipolar diffusion for $N = 2^{6-10}$ particles, using the M6 quintic kernel.  \emph{Bottom}: The cumulative root-mean-square error.}\label{fNWnoambi} 
\end{figure}

\begin{figure}
\begin{center}
\includegraphics[width=1.0\columnwidth]{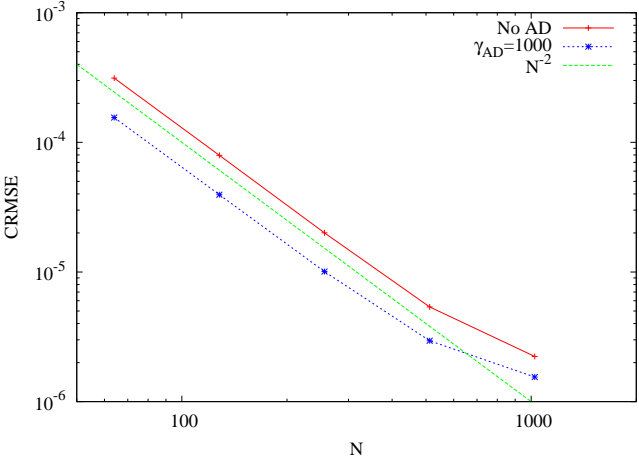}
\end{center}
\caption{The cumulative root-mean-square error at $t = 5$ as a function of resolution for the no ambipolar diffusion case (red) and the ambipolar diffusion case with $\gamma_\text{AD} = 1000$ (blue), along with a reference line $\propto N^{-2}$ (green).}\label{fNWcrmse} 
\end{figure}

Prior to implementing ambipolar diffusion in our test, we run a baseline test without ambipolar diffusion (that is, we turn off all ambipolar diffusion components in our code; analytically, this correspond to $\omega_\text{I} \equiv 0$).  We test $N = 2^{6-10}$ particles, and present time evolution of $<B_z^2>^{1/2}$ in Figure~\ref{fNWnoambi}.
We see good agreement between the analytical and numerical results, thus providing a useful verification that the code works correctly.  
Since the relative error~$\rightarrow \infty$ as $<B_z^2>^{1/2}\rightarrow 0$, a more useful error analysis is the cumulative root-mean-square error (CRMSE):
\begin{align}
&\text{CRMSE}(t) \equiv \\ \notag
&\sqrt{ \frac{1}{n}\sum_{i=1}^n \left\{\left[<B_z^2>^{1/2}(t_i)\right]_\text{analytical} -  \left[<B_z^2>^{1/2}(t_i)\right]_\text{numerical} \right\}^2 },
\end{align}
where there are $n$ outputs to time $t_n$ (inclusive), $t_i$ is the time of each output, and there is a constant $\text{d}t = 0.01$ between outputs.  As expected, this value continues to grow with time; the leading cause for the growth of this value is the slight difference between the analytical and numerical periods.  However, the CRMSE decreases with increasing resolution, indicating that increasing resolution allows for convergence to the analytical result. The red line in Figure~\ref{fNWcrmse} shows the CRMSE at $t = 5$ for each resolution, and we see that the slope is $\propto N^{-2}$, indicating a second order convergence. The CRMSE is not dependent on timestep for this test, indicating that the errors are dominated by the spatial terms.

We note that \citet{CKW09} perform their test using $v_0~=~0.1v_\text{A}$ rather than $0.01v_\text{A}$ used here.  Since the magnetic and velocity fields are dependent on one another, the oscillation excited in $B_z$ will excite an oscillation in $v_x$, which in turn will modify the oscillation of $B_z$.  Although this result is expected, it will yield a result that will differ from the analytical results used here, which were derived under the assumption of no non-linear coupling of waves.  Similar results are reached using the larger $v_0$, however the errors are noticeably worse and potentially misleading (i.e. for low resolution, the non-linear coupling is not well-defined, thus the errors remain small; for high resolution, the non-linear coupling is well defined and will ultimately converge, however, the errors are noticeably worse than the low resolution).  

\begin{figure}
\begin{center}
\includegraphics[width=1.0\columnwidth]{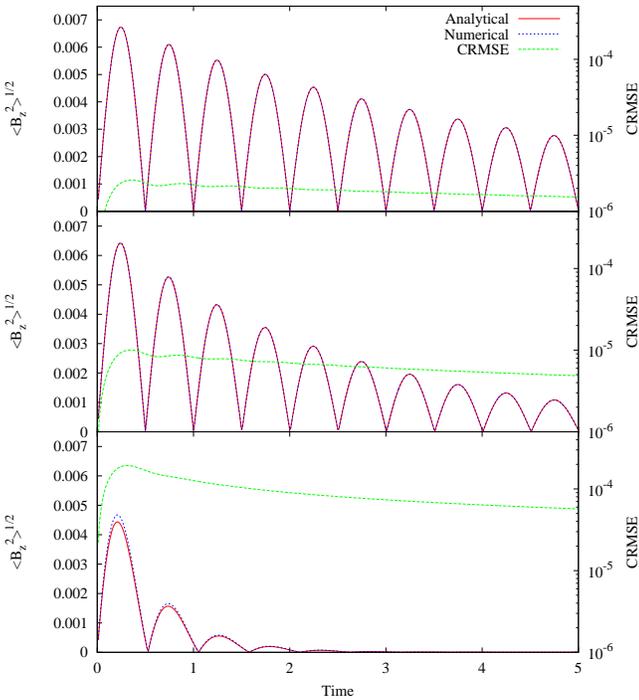}
\end{center}
\caption{Time evolution of the spatially averaged root-mean-square magnetic field in the $z$-direction to test the decay of Alfv{\'e}n waves in the strong coupling approximation.  Each panel includes the numerical result using $N = 1024$ particles (blue), the analytical result (red) and the cumulative root-mean-square error (green) for $\gamma_\text{AD} = $1000 (top), 500 (middle) and 100 (bottom).}\label{fNWgamma3} 
\end{figure}

For the remainder of our analysis we include ambipolar diffusion.  Figure \ref{fNWgamma3} shows the time evolution of $<B_z^2>^{1/2}$ for $\gamma_\text{AD} = $1000, 500 and 100 using $N = 1024$ particles.
We see agreement between the numerical and analytical results, with the CRMSE $\lesssim10^{-4}$ in all cases; the maximum CRMSE for $\gamma_\text{AD} = $100 is less than $\sim2$\% of the maximum $<B_z^2>^{1/2}$ amplitude, and is less than 0.03\% for $\gamma_\text{AD} = $1000.  The wave damps faster for decreasing $\gamma_\text{AD}$, so we are adding error in $<B_z^2>^{1/2}$ slower than $1/\sqrt{n}$ for increasing $t_n$, and the CRMSE decreases with time.  Since ambipolar diffusion damps the wave, the non-linear wave couplings are not as strong as compared to the no ambipolar diffusion case.  Thus, using the higher $v_0$ in this case yields similar errors to the case presented here.

For the case without ambipolar diffusion and the cases with $\gamma_\text{AD} = 1000$ and 500, the timestep is Courant limited.  For $\gamma_\text{AD} = 100$, the timestep is limited by ambipolar diffusion and is $\sim$4.4 times smaller than the Courant timestep.  Thus the impact of $\gamma_\text{AD}$ is apparent even in a simple test like this.  

\begin{figure}
\begin{center}
\includegraphics[width=1.0\columnwidth]{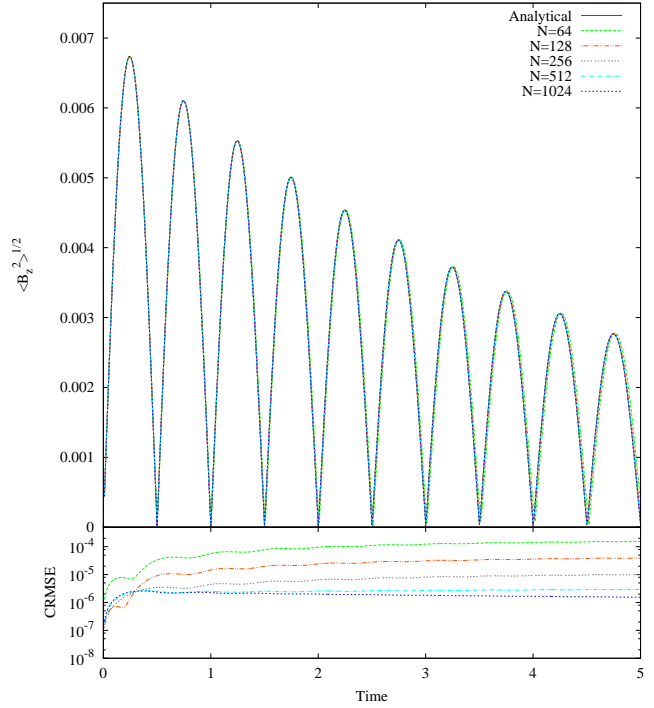}
\end{center}
\caption{\emph{Top}: Time evolution of the spatially averaged root-mean-square magnetic field in the $z$-direction to test the decay of Alfv{\'e}n waves in the strong coupling approximation; we test $N = 2^{6-10}$ particles for $\gamma_\text{AD} = 1000$ and the M$_{6}$ quintic kernel.  \emph{Bottom}: The cumulative root-mean-square error.}\label{fNWresolution} 
\end{figure}

We next present the results of a resolution test.  In Figure~\ref{fNWresolution}, we plot the time evolution of $<B_z^2>^{1/2}$ for $\gamma_\text{AD} = 1000$ using $N = 2^{6-10}$ particles.
Similar to the no ambipolar diffusion case, the results get better with resolution.  The blue line in Figure~\ref{fNWcrmse} indicates that, even with the inclusion of ambipolar diffusion, we obtain second order convergence.  

\begin{figure}
\begin{center}
\includegraphics[width=1.0\columnwidth]{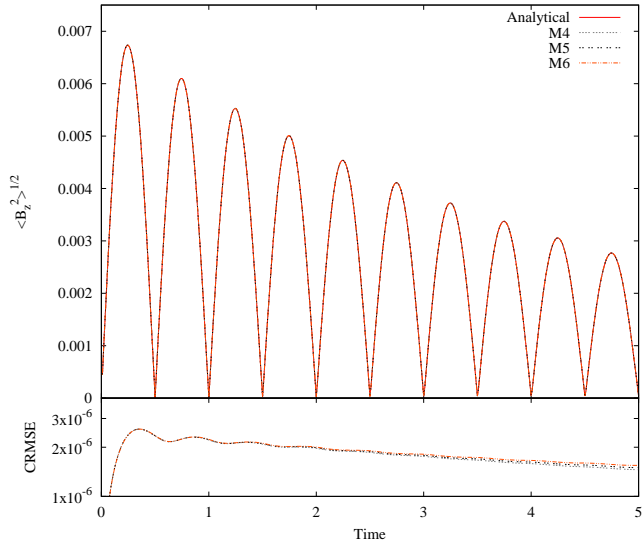}
\end{center}
\caption{\emph{Top}: Time evolution of the spatially averaged root-mean-square magnetic field in the $z$-direction to test the decay of Alfv{\'e}n waves in the strong coupling approximation; we test the M$_{4}$ cubic, M$_{5}$ quartic and M$_{6}$ quintic smoothing kernels using $N = 1024$ particles and $\gamma_\text{AD} = 1000$.  \emph{Bottom}: The cumulative root-mean-square error.}\label{fNWkernel} 
\end{figure}

To test the dependence on smoothing kernels, we run the $\gamma_\text{AD} = 1000$ run with $N = 1024$ using the M$_{4}$ cubic, M$_{5}$ quartic and M$_{6}$ quintic smoothing kernels; see Figure \ref{fNWkernel}.
The CRMSE increases as the kernel moves from M$_{4}$ to M$_{6}$, which is reasonable since M$_{6}$ smoothes over a greater number of particles than the M$_{4}$ kernel.  At this resolution, the CRMSE at $t = 5$ for all three kernels are equal within 6.1\% .  If we decrease resolution to $N = 512$ (128), the relationship between CRMSE and the kernels is the same, but the separation increases such that at $t = 5$, the three CRMSE values are equal within 26.6\% (39.2\%).   Although the cubic smoothing kernel yields lower CRMSE values, the quintic kernel yields stable results for a longer time given ideal initial conditions (i.e. particles evenly spaced on a 1D lattice).  This stability issue has previously been studied on a cubic lattice \citep[e.g.][]{morris96,price12}, thus we will not investigate it further.  For this analysis, we have chosen stability over CRMSE values, hence our choice of the quintic smoothing kernel.

\begin{figure}
\begin{center}
\includegraphics[width=1.0\columnwidth]{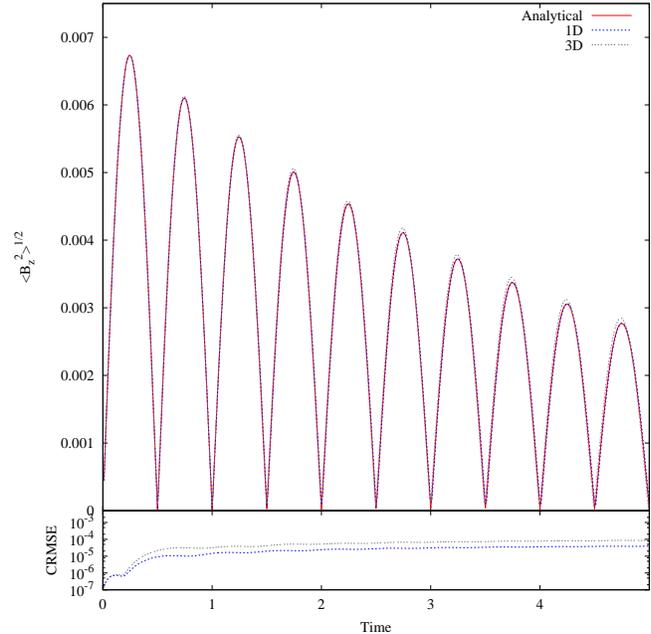}
\end{center}
\caption{\emph{Top}: Time evolution of the spatially averaged root-mean-square magnetic field in the $z$-direction to test the decay of Alfv{\'e}n waves in the strong coupling approximation; we compare the 3D model to the 1D model and to the analytic solution (red line) using $\gamma_\text{AD} = 1000$.  The 1D model uses the M$_{6}$ quintic kernel and $N_\text{1D} = 128$ particles; the 3D model uses the $\mathcal{C}^4$ Wendland kernel and $N_\text{1D} = 126$ particles.  \emph{Bottom}: The cumulative root-mean-square error.}\label{fNW1d3d} 
\end{figure}

Finally, we test our ambipolar diffusion algorithm in the fully 3D SPMHD code {\sc Phantom} \citep{pricefederrath10,lodatoprice10}.  Given that this code uses different dissipation, magnetic cleaning and integration algorithms than our 1D code, we do not expect identical results when comparing the results at similar resolutions.  Moreover, the stability of a cubic lattice is a well-known issue in SPH \citep[e.g.][]{morris96,price12}, thus the particles are initialised on a close-packed lattice; further, the $\mathcal{C}^4$ Wendland kernel is used to maintain stability over five periods.   In {\sc Phantom}, we set $\gamma_\text{AD} = 1000$, $N_\text{3D} \equiv N_\text{1D}^3$ with $N_\text{1D}$ = 126, $L_x = 1$ (as in the 1D case), $L_y = \frac{\sqrt{3}}{2}L_x$ and $L_z = \frac{\sqrt{6}}{3}L_x$, and present the comparison in Figure~\ref{fNW1d3d}.

The CRMSE at $t = 5$ is $\sim$2 times higher for the 3D code over the 1D code, thus  we can be confident that {\sc Phantom} can successfully model ambipolar diffusion.  

For completeness, when the 3D test was run on a cubic lattice using the M$_6$ quintic kernel, the particles fell off the lattice after one period for $N_\text{1D} = 128$.  Just prior to the particles falling off of the lattice, the CRMSE for the 1D case was $\sim$56\% larger than that for the 3D case.  Although the CRMSE values are smaller than the values presented here, we chose to present the results using the Wendland kernel since it produced a stable result over at least five periods.  

\subsection{Oblique C-shock}

The only other known solution for benchmarking ambipolar diffusion codes is the C-shock.  The setup of this requires that an isothermal, neutral gas in a magnetic field be reflected off of a wall.  If the gas velocity exceeds the sound speed but not the Alfv{\'e}n speed, then an Alfv{\'e}n wave moving through the ions will drag the neutrals into the post-shock region without creating a discontinuity, creating the C-shock \citep{Draine80}.  A semi-analytic solution to this can be calculated by taking the steady-state MHD equations (i.e. setting $\text{d}t=0$ and $\text{d}x = \text{d}y = \text{d}z = 0$ in \eqref{CErho}, \eqref{CEv} and \eqref{CEb}), reducing them to a single ordinary differential equation and then numerically solving the ODE (e.g. \citealt{Wardle1991, MNKW95, DP08, CKW09}).  We explicitly note that, unlike (e.g.) the Sod shock tube problem, this solution describes only the steady-state shock and not the entire domain of the model.  Thus, any additional shocks created as a result of reflecting off of the wall are not described by this semi-analytical solution.  

The simplest numerical implementation of a C-shock is to initialise a velocity towards a reflective boundary.  However, to avoid implementing a reflective boundary condition in SPH, we instead initialise two inflows that will meet at the centre of the domain.  The magnetic field is continuous across the entire domain, which is equivalent to using a continuous boundary condition for the magnetic field, as used in \citet{MNKW95}.  Our initial conditions are $\rho = 1$, $c_\text{s} = 0.1$, $\bm{B} = B_0\bm{\hat{x}} + B_0\bm{\hat{y}}$ with $B_0 = \frac{1}{\sqrt{2}}$, $\bm{v} = v_0\bm{\hat{x}}$ with $v_0 = -4.45$ (4.45) for $x > 0$ ($x < 0$), $\rho_\text{i} = 10^{-5}$ and $\gamma_\text{AD} = 1$.  Our domain extends for $-40L_\text{AD} < x < 40L_\text{AD}$, with an initial particle separation of $0.2L_\text{AD}$.  Analytical results predict a post-shock density approximately eight times higher, thus yield a post-shock separation of $0.025L_\text{AD}$, and hence the shock resolution will be between these two values\footnote{The constant resolution in the grid simulations presented in \citealt{MNKW95} is 0.2 and $0.1L_\text{AD}$, and in \citet{CKW09} the resolution is $20L_\text{AD}/128 \approx 0.156L_\text{AD}$.}.  Due to the reflective nature of our results, we will only present the results for the domain $x > 0$.  As with the wave damping test, we also run a baseline test without ambipolar diffusion.

\begin{figure}
\begin{center}
\includegraphics[width=1.0\columnwidth]{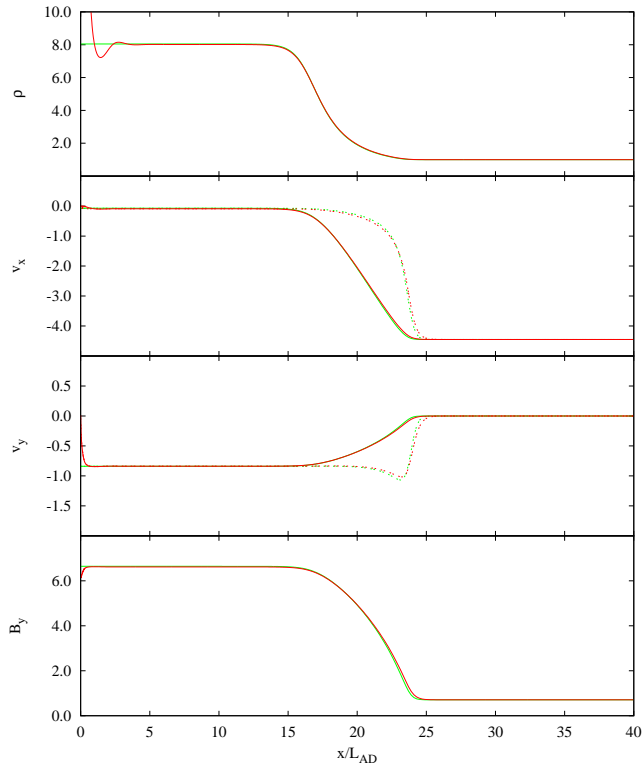}
\end{center}
\caption{Results of the C-shock test after $t = 4\tau_\text{AD}$, using  $\rho_\text{i} = 10^{-5}$ and $\gamma_\text{AD} = 1$.  The red lines are the numerical results and the green lines are the semi-analytical results.  \emph{Top to Bottom}: Neutral gas density; neutral (solid) and ion (dotted) velocity in the $x$-direction; neutral (solid) and ion (dotted) velocity in the $y$-direction; magnetic field strength in the $y$-direction.}\label{fNC1} 
\end{figure}

Figure \ref{fNC1} shows the neutral density, neutral and ion velocities and the $y$-component of the magnetic field after $t = 4\tau_\text{AD}$.
For the C-shock structure ($10\lesssim x/L_\text{AD} \lesssim 30$), we find agreement to within $\sim$4\% between the numerical and semi-analytical results.  There is an increase in the relative error at the base of the shock, but this can be attributed to our artificial viscosity and resistivity algorithms, and is not related to our implementation of ambipolar diffusion.  A brief study shows that the numerical results converge for increasing resolution, but the runtime is severely hampered due to the timestepping limitations (recall Eq. \ref{Idt}) and the short smoothing lengths near $x = 0$.  Although the shortest smoothing length is larger for the case with ambipolar diffusion, the quadratic dependence on it results in a timestep that is 30-40 times lower than for the case without ambipolar diffusion (whose timestep dependence on smoothing length is linear).

For completeness, we note that the C-shock is not the complete solution to the system -- a second shock exists near the origin.  This shock appears in the numerical solution presented in \citet{CKW09} (although it is not discussed), and is not shown in \citet{MNKW95} since the boundary is removed from the plot. These authors cite ``wall heating'' as the source of this discontinuity, but this is clearly not the case since the problem is isothermal.  This second shock is also present in the case without ambipolar diffusion, and this profile is similar to the solution of the MHD shock given in fig. 2a of \citet{RJ95}\footnote{The extra density discontinuity in fig. 2a of \citet{RJ95} is a result of their initial discontinuity in density; this discontinuity is absent in our results since our initial density is constant everywhere.}.  This is expected since both tests are initialised with inflow velocities.  We ran this test without ambipolar diffusion and with $c_\text{s} = 1.0$ using both our simple SPMHD code and {\sc Athena} \citep{Athena}.  Both codes produced the shock near the origin, indicating that it is a real feature.   As expected, this shock is also smoothed by ambipolar diffusion, similar to how the first shock is smoothed to create the C-shock profile. 

\section{Summary}
\label{sec:summary}
 We have described a simple implementation of ambipolar diffusion suitable for SPMHD codes.  The same algorithm can be easily extended to handle Ohmic resistivity and to the Hall effect.  Our derivation assumed the strong coupling approximation ($\rho \sim \rho_\text{n}$ and $\rho_\text{ion} \ll \rho_\text{n}$) and thus we can use a single fluid approach.  We have shown that this method conserves energy, and the contribution to the energy equation is always positive definite, as required.  We have tested this implementation in both a simple 1D SPMHD code and the fully 3D code {\sc Phantom}.  Our results are as follows:
\begin{enumerate}
\item For the wave damping test, our numerical results agreed with the analytical results.  For the three cases we studied, the cumulative root-mean-square error remained less than 2\% of the maximum $<B_z^2>^{1/2}$ amplitude over five periods, with larger values of the collisional coupling constant, $\gamma_\text{AD}$, yielding smaller errors (i.e. 0.03\% error for  $\gamma_\text{AD}$=1000 compared to 2\% for  $\gamma_\text{AD}$=100).
\item Our implementation is robust to resolution and kernel tests.  For increasing resolution, the convergence is second order.  As the smoothing kernel is switched from the quintic to the quartic to the cubic, the cumulative root-mean-square error decreases, as to be expected since the kernel is smoothing over a shorter distance.  
\item The analytical results can be reproduced using a fully 3D SPMHD code.  The cumulative root-mean-square error is lower than for the 1D code.
\item For the oblique C-shock test, our numerical results agreed with the semi-analytical results typically within $\sim$4\%.  Although the relative error was larger than 4\% at the base of the shock, this can be attributed to the artificial viscosity and resistivity, and not the implementation of ambipolar diffusion.
\end{enumerate}
With the inclusion of ambipolar diffusion in 3D SPMHD codes, we are now in a position to determine the effect that it has on physical processes, such as the collapse of molecular clouds to form stars. 

\section*{Acknowledgments}
We thank the anonymous referee for helpful comments that improved the clarity and overall quality of the manuscript.  We gratefully acknowledge funding via Australian Research Council Discovery Grant DP130102078. DJP is supported by an ARC Future Fellowship (FT130100034).

\bibliography{Wbib,Pbib}

\label{lastpage}
\end{document}